\documentclass[pre,aps,twocolumn,showpacs]{revtex4}
\usepackage{amsmath,amssymb,graphicx,cases}
\usepackage{epsfig}
\usepackage{textcomp}
\usepackage{color}

\begin{document}
\title{Large-displacement statistics of the rightmost particle of the  one-dimensional branching Brownian motion}
\author{Bernard Derrida}
\email{derrida@lps.ens.fr}
\affiliation{Coll\`{e}ge de France
 11 Place Marcelin Berthelot, 75005 Paris
\\
and LPS,  \'{E}cole Normale Sup\'{e}rieure,
24 rue Lhomond, 75005 Paris, France}
\author{Baruch Meerson}
\email{meerson@mail.huji.ac.il}
\affiliation{Racah Institute of Physics, Hebrew University of
Jerusalem, Jerusalem 91904, Israel}
\author{Pavel V. Sasorov}
\email{pavel.sasorov@gmail.com}
\affiliation{Keldysh Institute of Applied Mathematics, Moscow 125047, Russia}

%02.50.Ga   Markov processes
%05.40.-a – Fluctuation phenomena, random processes, noise, and Brownian motion
\begin{abstract}
Consider a one-dimensional branching Brownian motion, and rescale the coordinate and time so that the rates of branching and diffusion are both equal to $1$.
If $X_1(t)$ is the position of the rightmost particle of the branching Brownian motion  at time $t$, the empirical velocity $c$ of this rightmost particle is defined as $c=X_1(t)/t$. Using the Fisher-KPP equation, we evaluate the probability distribution ${\mathcal P(c,t)}$ of this empirical velocity $c$ in the  long time $t$  limit for $c > 2$. It was already known
that, for a single seed particle,   ${\mathcal P(c,t)} \sim \exp \,[-(c^2/4-1)t]$ up to a prefactor that can depend on $c$ and $t$. Here we show how to determine this prefactor. The result can be easily generalized to the  case of multiple seed particles and to branching random walks associated to other traveling wave equations.

\end{abstract}
\maketitle

\section{Statement of the problem}
\label{intr}

Branching Brownian motion unites two classical con\-tinuous-time Markov processes: the random branching and
the Brownian motion, or the Wiener process. The branching Brownian motion was extensively studied in the past \cite{McKean,Bramson}, and it continues to attract much attention among physicists \cite{BD2009,Mueller,Ramola1}.  In the basic version of the branching Brownian motion that we consider here each particle  can give birth, in a small
time interval $\Delta t$,  to another particle
with probability $\lambda \Delta t$. It can also perform, with the complementary probability $(1-\lambda \Delta t)$, Brownian motion with diffusion coefficient $D$. By rescaling time and the coordinate we can set $\lambda=D=1$.
Here we consider the branching Brownian motion in one dimension. Suppose that the process starts at $t=0$ either with a single particle at the origin, or with many particles
distributed in space so that the rightmost particle is at the origin.  Our goal is to evaluate the long-time probability that the position $X_1(t)$ of the rightmost particle at time $t$ corresponds to a given velocity $c=X_1(t)/t$. We are interested in the case of $c>2$, including arbitrary large $c$.

Our approach is  based on a remarkable result of McKean \cite{McKean,Bramson} that we now state.  Let
\begin{equation}
\label{P(x,t)}
P(x,t) = \mbox{Prob}\{X_1(t)>x\}
\end{equation}
be the probability of observing the rightmost particle at  the right of a given point $x$.
If the process starts with a single seed particle at $t=0$, this probability $P(x,t)=u(x,t)$ obeys the equation:
\begin{equation}
\partial_t u =u-u^2 +\partial_x^2 u
\label{FKPP}
\end{equation}
with the initial condition
\begin{equation}
u(x,t=0)=\theta(-x),
\label{step}
\end{equation}
where $\theta(\dots)$ is the step function.  Equation (\ref{FKPP}) is the celebrated Fisher-Kolmogorov-Petrovsky-Piscounov (FKPP) equation ~\cite{Fisher,KPP}. It appears in
mathematical genetics and dynamics of populations   \cite{Fisher,KPP,Murray}.  It is also encountered
in chemical kinetics \cite{Douglas},  theory of disordered systems \cite{Spohn},
extreme value statistics \cite{MajKr},
and high-energy physics \cite{particlephys}. With an initial condition like Eq.~(\ref{step}),
the FKPP equation describes the invasion of an unstable phase $u=0$ by a stable phase $u=1$.

The possible invasion front solutions of Eq.~(\ref{FKPP}) are described by a family of traveling-front solutions (TFSs) of Eq.~(\ref{FKPP}), parameterized by their velocity $v$:
$u(x,t)=\Psi_{v}(\xi)$, where $\xi=x-vt$.
$\Psi_{v}(\xi)$ obeys the  equation
\begin{equation}\label{MFeq}
   \Psi_{v}^{\prime\prime}+v \Psi_{v}^{\prime}+\Psi_{v}-\Psi_{v}^2=0
\end{equation}
(where the prime denotes the $\xi$-derivative) and the boundary conditions $\Psi_v(-\infty)=1$ and $\Psi_v(+\infty)=0$.
An admissible invasion front could have any velocity from the interval
$v_*\leq v<\infty$, where $v_*=2$.

It is known that for  a steep initial condition like Eq.~(\ref{step}), the solution of Eq.~(\ref{FKPP}) converges at long times to the limiting TFS, $\Psi_{2}$, of Eq.~(\ref{MFeq}) with the velocity $v_*=2$ \cite{Bramson,KPP,Saarloos} up to a non-trivial shift in the position of the front  \cite{Bramson,Saarloos,Derridashift,EbertSaarloos}.

In the present work we  want to study
the probability distribution that the position $X_1(t)$ of the rightmost particle at a long time $t$ has traveled with an empirical  velocity $c=X_1(t)/t$ such that $2<c<\infty$. To achieve this goal, we will have to go beyond previous asymptotic results on the behavior of $u(x,t)$ in the leading edge of the evolving front.

Before we delve into this issue, let us note that   the
solution of (\ref{FKPP}) and (\ref{step})  allows one to obtain the probability distribution of the rightmost particle when initially, at $t=0$,  there are $M$ seed particles at the points $x_m$ ($m=1,2, \dots , M$), so that $x_{m+1}\leq x_m$. Without loss of generality, we can assume that the rightmost seed particle is initially at $x=0$, so that $x_1=0$. Since all particles in the branching Brownian motion are independent, the probability (\ref{P(x,t)})
can be expressed as
\begin{equation}
P(x,t)=1-\prod\limits_{m=1}^M \left[1-u(x-x_m,t)\right],
\label{E040}
\end{equation}
where $u(x,t)$ is the solution of the problem (\ref{FKPP}) and (\ref{step}).
Equation~(\ref{E040}) allows one to calculate the probability distribution
${\mathcal P}(c,t)$ of the empirical velocity $c=X_1(t)/t$
  of the rightmost particle
\begin{equation}
{\mathcal P}(c,t)=-\frac{d}{dc} P(ct,t).
\label{E050}
\end{equation}
We will evaluate ${\mathcal P}(c,t)$ for a single seed particle, for several seed particles and in some cases for  infinitely many seed particles. Up to pre-exponential factors, the large-time behavior of
${\mathcal P}(c,t)$
is  the same in all cases for $c>2$,
\begin{equation}\label{result}
  \frac{\ln {\mathcal P}(c,t)}{t} \simeq 1-\frac{c^2}{4}, \quad t\gg 1,
\end{equation}
but the  pre-exponential factors are different.
For a single seed particle,   Eq.~(\ref{result}) has been known for a long time \cite{Rouault}. Here we show how to determine the pre-exponential factors.

The remainder of the paper is structured as follows.
Section \ref{single}
deals with a long-time leading-edge asymptotic of the solution of the problem described by Eqs.~(\ref{FKPP}) and (\ref{step}).
We establish an applicability criterion of a long-time asymptotic found earlier. We derive in Sec. \ref{BD} the  long-time
leading-edge asymptotic which  holds at an arbitrarily large distance to the right from the instantaneous front position.

In Sec. \ref{probdens} we apply our asymptotics to  evaluate the probability distribution ${\mathcal P(c,t)}$
of the empirical velocity $c=X_1(t)/t>2$ of the rightmost particle
for a single seed particle, and for multiple seed particles. We briefly discuss our results in Sec. \ref{discussion}. The generalization to other traveling wave equations is discussed in Appendix B.

\section{Long-time behavior of the leading edge of the FKPP front}
\label{single}

Let us recap some important properties of the solution of the problem (\ref{FKPP}) and (\ref{step})~\cite{Bramson,Derridashift,EbertSaarloos,Saarloos}. At long times,  $t\gg 1$, the FKPP front position $x_f(t)$ is  given by the asymptotic formula
\begin{equation}\label{xf}
x_f(t)=2t-\frac{3}{2}\ln t + A -3\sqrt{\frac{\pi}{t}}+{\mathcal O}\left(\frac{1}{t}\right) \,,
\end{equation}
where $A$ is a constant.  At fixed $v$, the TFS $\Psi_v(\xi)$ is unique only up to translations in $\xi$. As a result, the constant $A$ can be defined uniquely only if this translation freedom is eliminated: for example, by fixing $\Psi_v(0)$.  We will fix $\Psi_v(0)=1/2$; in this case  a numerical solution of the problem (\ref{FKPP}) and (\ref{step}) yields $A\simeq -1.8$.

At $t\gg 1$, and at relatively small distances  from the front,  $|x-x_f(t)|\ll\sqrt{t}$, the solution of the problem ~(\ref{FKPP}) and (\ref{step}) can be approximated as
\begin{equation}\label{asymp}
u(x,t)=\Psi_2(x-x_f(t))+ \dots\,.
\end{equation}
where dots denote a small correction that vanishes at $t\to \infty$. We will also need for our purposes the
large-$\xi$ asymptotic of the function $\Psi_2(\xi)$:
\begin{equation}
 \label{TFS}
\Psi_2(\xi \gg 1)=\gamma (\xi-\xi_0 +o(1))\, e^{\xi_0-\xi}\,.
\end{equation}
The constants $\gamma$ and $\xi_0$ can be found numerically. $\xi_0$ depends on the choice of $\Psi_2(0)$, while $\gamma$ is independent of it.  A numerical solution of Eq.~(\ref{MFeq}) gives $\gamma \simeq 0.142$ and, with out choice of  $\Psi_2(0)=1/2$, $\xi_0\simeq 3.22$.

In  \cite{Derridashift}
Brunet and Derrida
(see also Ref. \cite{EbertSaarloos}) derived a long-time leading-edge asymptotic which holds in a region
ahead of the instantaneous front position (\ref{xf}). The size of this region is expanding with time, see below. In this region the  solution $u(x,t)$ of Eqs.~(\ref{FKPP}) and (\ref{step})   can be approximated by the solution of the linear equation
\begin{equation}
\partial_tu= u+ \partial_x^2u .
\label{B040A}
\end{equation}
 Brunet and Derrida obtained the following asymptotic \cite{equationnumbers}:
\begin{equation}
u(x,t) \simeq  \gamma z \sqrt{t} \exp\left(-z\sqrt{t}-\frac{z^2}{4}\right),
\label{A010}
\end{equation}
where
\begin{equation}\label{z}
z=\frac{x-2t+\frac{3}{2} \ln t- \Delta }{\sqrt{t}},
\end{equation}
and $z$ is assumed to be (positive and) of order $1$. The constant $\Delta =A+\xi_0 \simeq 1.42$ is independent of the specific choice of $\Psi_2(0)$.

What are the applicability conditions of the long-time asymptotic~(\ref{A010})? One condition already mentioned is
\begin{equation}\label{leftinequality}
1 \ll x-2t+\frac{3}{2}\ln t ,
\end{equation}
which justifies the linearization of the FKPP equation in the leading edge of the front \cite{Derridashift}.   An additional condition limits the applicability of
Eq.(\ref{A010}) at large distances
ahead of the front. A sufficient applicability condition  $z = {\cal O}(1)$, i.e. $x-2t+(3/2) \ln t   = {\cal O}(\sqrt{t})$ has been known since the work of Brunet and Derrida \cite{Derridashift}.  In the next subsection we will relax this condition to the following form:
\begin{equation}
\label{rightinequality}
x-2t+\frac{3}{2}\ln t \ll \frac{t}{\ln t}\,.
\end{equation}
In view of  the strong inequalities (\ref{leftinequality}) and (\ref{rightinequality}), the asymptotic (\ref{A010}) can be rewritten as
\begin{equation}
u(x,t)\simeq \alpha\, \frac{x-2t+\frac{3}{2}\ln t}{t^{3/2}}\, e^{t-\frac{x^2}{4t}}\, ,
\label{A140}
\end{equation}
where $\alpha=\gamma \,e^\Delta\simeq 0.59$ is independent of the specific choice of
$\Psi_2(0)$.

The improved condition (\ref{rightinequality}) is still too restrictive for our purpose of dealing with $c= x/t>2$
at arbitrary large $t$ and, in particular, dealing with an arbitrary large $c$. Therefore, in Sec. \ref{BD} we will derive a generalization of Eq.~(\ref{A140})
which, in the long time limit,   holds at an arbitrarily large distance to the right from the instantaneous front position.
As to be expected from  Eq.~(\ref{result}), the exponential dependence in Eq.~(\ref{A140}) persists in the new asymptotic,
\begin{equation}\label{onlyexp}
    u(x,t) \sim e^{t-\frac{x^2}{4t}} ,
\end{equation}
but the prefactor, which we will find, in general depends on $c$ in a different way.

%\vspace{0.5cm}

\subsection{Derivation of the condition (\protect\ref{rightinequality})}
\label{DER}

In  \cite{Derridashift}  the linearized equation (\ref{B040A})  was solved at $t\gg 1$
 and $z = {\cal O}( 1)$
by making an ansatz which, for the initial condition (\ref{step}) of the complete nonlinear problem,
has the following form:
\begin{equation}\label{Dansatz}
u(x,t)=\sqrt{t}\, G(z)\, e^{-z\sqrt{t}}\,,
\end{equation}
where $z$ is defined in Eq.~(\ref{z}). The asymptotic (\ref{A140}) [or~(\ref{A010})] corresponds to
\begin{equation}\label{G}
G(z)={\gamma} z e^{-z^2/4}.
\end{equation}
Here, in order to estimate the range of validity of (\ref{G}), we consider a more general ansatz
\begin{equation}
u(x,t)=\sqrt{t}\, g(z,t)\, e^{-z\sqrt{t}}
\label{A020}
\end{equation}
where $g(z,t)$ depends both on $z$ and on time. This ansatz makes it possible to track the approach of the solution of the linear equation (\ref{B040A}) to the scaling form (\ref{Dansatz}) at large times. Plugging Eq.~(\ref{A020}) into Eq.~(\ref{B040A}), one obtains an exact equation for the function $g(z,t)$:
\begin{equation}
\partial_z^2g+\frac{z}{2}\, \partial_z g+g -t\, \partial_t g =\frac{3}{2\sqrt{t}}\, \partial_z g\, .
\label{A040}
\end{equation}
At $z={\mathcal O}(1)$, the right-hand side  of Eq.~(\ref{A040}) becomes negligibly small at $t\to \infty$,  and this is what was used
in the derivation of Eq.~(\ref{A140}) in Ref. \cite{Derridashift}. Extending their approach, we can seek a perturbative solution at $t\gg 1$:
\begin{equation}
g(z,t)=g_0(z)+\frac{g_1(z)}{\sqrt{t}}+\dots+\frac{g_n(z)}{(\sqrt{t})^n}+\dots\,,
\label{A030}
\end{equation}
where $g_1(z)/\sqrt{t}\ll g_0(z)$, etc. In the zeroth order of the perturbation expansion we  get as in  \cite{Derridashift}
\begin{equation}\label{A050}
g_0^{\prime\prime}+\frac{z}{2}\, g_0^{\prime}+g_0 = 0 \ .
\end{equation}
Its proper solution, obtained by matching with the traveling wave solution (\ref{TFS}) of the Fisher equation [Eq.~(\ref{MFeq}) with $ v=2$], is given by Eq.~(\ref{G}).

In the first order of the perturbation expansion we obtain a forced linear equation for $g_1(z)$:
\begin{eqnarray}\label{A070}
g_1^{\prime\prime}+\frac{z}{2}\, g_1^{\prime}+\frac{3}{2}\, g_1 &= &\frac{3}{2}g_0^{\prime}\nonumber \\
&=&\frac{3{\gamma}}{2}\, \left(1-\frac{z^2}{2}\right)\, e^{-z^2/4}\, ,
\end{eqnarray}
The ansatz $g_1(z)=h_1(z) e^{-z^2/4}$ reduces this equation to
\begin{equation}\label{reduced}
    h_1^{\prime\prime}-\frac{z}{2} h_1^{\prime}+h_1 = \frac{3{\gamma}}{2}\left(1-\frac{z^2}{2}\right)\,.
\end{equation}
This equation can be solved  analytically, but the complete solution is quite cumbersome. Fortunately, we only need the  $z\gg1$ asymptotic of the solution for the purpose of derivation of the applicability criterion
(\ref{rightinequality}).
Importantly, the forcing term of Eq.~(\ref{reduced}) coincides, up to a constant multiplier, with one independent solution of the homogeneous equation.
 The fastest growing term of $h_1(z)$ at large $z$ can be found by dropping the second derivative term on the left-hand side and neglecting $1$ compared with $z^2/2$ on the right hand side. The resulting first order equation,
\begin{equation}
z h_1^{\prime} - 2h_1 =\frac{3\gamma}{2}z^2\,,
\label{D080}
\end{equation}
has the following general solution:
$$
h_1(z)=C z^2+\frac{3\gamma}{2}\,z^2 \ln z,
$$
where $C$ is an arbitrary constant. The leading-order asymptotic of $h_1(z)$ at $z\gg 1$,
$h_1(z) \simeq(3\gamma/2)\,z^2 \ln z$, is dominated by the forced solution.  Correspondingly, the leading-order asymptotic of $g_1(z)$ is \cite{cumbersome}
\begin{equation}
g_1(z\gg 1)= \frac{3\gamma}{2} e^{-z^2/4} z^2\ln z \,.
\label{D090}
\end{equation}
That $g_1(z) \gg g_0(z)=G(z) $ for large $z$  in (\ref{A030}) suggests that the limits $t \to \infty$ and $z \to \infty$ do not commute
and  that  the asymptotics (\ref{A010}) or  the ansatz (\ref{Dansatz})  are only valid when one takes the limit $t \to \infty$ first.

If we  demand  that $g_1(z)/\sqrt{t}$ be small compared with $g_0(z)$ in (\ref{A030}) we get
the inequality
\begin{equation}\label{critright}
\frac{z\ln z}{\sqrt{t}} \ll 1.
\end{equation}
Using Eq.~(\ref{z}) for $z$ and replacing, with logarithmic accuracy, $z$ by $t$ under $\ln z$, we obtain the criterion (\ref{rightinequality}).

In the higher orders of the perturbation expansion (\ref{A030}) we obtain
\begin{equation}\label{rec}
g_n^{\prime\prime}+\frac{z}{2}g_n^{\prime}+\frac{n+2}{2}g_n=\frac{3}{2}g_{n-1}^{\prime}\, .
\end{equation}
Similarly to the case of $n=1$, we make the ansatz $g_n(z)=h_n(z) e^{-z^2/4}$. Solving the resulting equation for
$h_n(z)$, we obtain the $z\gg 1$ asymptotic
\begin{equation}\label{gnas}
g_n(z\gg 1) \simeq \left(\frac{3}{2}\right)^n \gamma z^{n+1} \ln^n z \,e^{-z^2/4}\,,
\end{equation}
which again yields the criterion (\ref{critright})
and therefore (\ref{rightinequality}).

 As we will show in subsection \ref{Sc-2}, the applicability of asymptotic (\ref{A010}) can be actually extended, see Eq.~(\ref{extended}).

%\vspace{0.5cm}
\section{The leading-edge asymptotic of the FKPP front at $(c-2)^2t\gg 1$}
\label{BD}

 We are interested in an asymptotic behavior of the leading edge of the FKKP front at long times $t$ in the region where $x/t=c=\mbox{const}>2$. As we will see, the  ``long times" imply $(c-2)^2 \,t \gg 1$.
Our starting point here is the exact expression
\begin{eqnarray}
  &&u(x,t)-u_{\text{linear}}(x,t)\nonumber \\
  && =-\int_0^t d\tau\int_{-\infty}^{\infty} d\chi\,\,\frac{e^{t-\tau-\frac{(x-\chi)^2}{4 (t-\tau )}}}{\sqrt{4\pi(t-\tau )}} \,
  u^2(\chi,\tau), \label{exact}
\end{eqnarray}
where
\begin{equation}\label{psilinear0}
u_{\text{linear}}(x,t) = \frac{e^t}{2}\,\text{erfc} \left(\frac{x}{\sqrt{4t}}\right).
\end{equation}
is the solution of the linear equation (\ref{B040A}) with the initial condition (\ref{step}).
To derive Eq.~(\ref{exact}) we formally treat Eq.~(\ref{FKPP}) as an inhomogeneous linear equation:
\begin{equation}\label{formalinear}
\partial_t u-\partial_x^2 u - u=-u^2.
\end{equation}
A formal solution of this equation with the initial condition~(\ref{step}) is obtained as convolution of the Green's function of the left-hand-side operator with the right hand side of the equation. This yields the term (\ref{psilinear0}) and the double integral in Eq.~(\ref{exact}).
Setting $x=ct$ in the right-hand side of   Eq.~(\ref{exact}),  we obtain
\begin{eqnarray}
    &&u(ct,t)-u_{\text{linear}}(ct,t) \nonumber \\
  && = e^{\left(1-\frac{c^2}{4}\right)\, t}
  \int\limits_0^t  d\tau \int\limits_{-\infty}^{\infty} d\chi\,  \frac{ u^2(\chi,\tau )}{\sqrt{4\pi (t-\tau )}} \nonumber \\
  && \times \,\exp{\left[-\frac{(4+c^2)}{4}\, \tau +\frac{c\chi}{2}-\frac{(\chi-c\tau)^2}{ 4(t-\tau )}\right]}.
  \label{C010}
\end{eqnarray}
At $c=x/t=\text{const}>2$ and $t\to\infty$
the integral is dominated by $\chi \sim \tau \ll t $, and
we can make three simplifications:  (i) neglect $\tau $ compared with $t$ under the square root,  (ii) neglect the term
$$
\frac{(\chi-c\tau)^2}{ 4(t-\tau )}
$$
in the exponent, and (iii) send  $t$ to $\infty$ in the upper limit of integration over $\chi$. We will check the validity of these assumptions a posteriori. After these simplifications, Eq.~(\ref{exact}) becomes, in the leading order in $1/t$,
\begin{equation}
  u(ct,t)-u_{\text{linear}}(ct,t) \simeq-\frac{e^{-t(\frac{c^2}{4}-1)}}{\sqrt{4 \pi t}}\,\phi(c),
  \label{formula0}
\end{equation}
where
\begin{equation}
\phi(c)=
\int_{0}^{\infty} \!\!d\tau
\int_{-\infty}^{\infty} \!\!d\chi\,
 e^{-\left(1+\frac{c^2}{4}\right) \tau +\frac{c \chi}{2}} u^2(\chi,\tau ).\label{formula}
\end{equation}
In the same leading order in $1/t$, Eq.~(\ref{psilinear0}) simplifies to
\begin{equation}\label{psilinear}
u_{\text{linear}}(ct,t) \simeq \frac{e^{-t(\frac{c^2}{4}-1)}}{c\sqrt{\pi t}},
\end{equation}
so we can rewrite Eq.~(\ref{formula0}) as
\begin{equation}
  u(ct,t)\simeq \frac{e^{-t(\frac{c^2}{4}-1)}}{\sqrt{4 \pi t}}\,\Phi(c), \label{formula1}
\end{equation}
where
\begin{equation}\label{Phiphi}
\Phi(c)=\frac{2}{c} -\phi(c)\, .
\end{equation}
To see that the integral in Eq.~(\ref{formula}) converges, and the iterated integrals over $\tau$ and over $\chi$ commute at any $c\geq 2$, let us separately consider the cases of $2\leq c<4$ and $c\geq 4$.
At $2\leq c<4$ and
large $\tau$, the
integrand as a function of $\chi$ has a narrow maximum at $\chi\simeq x_f(\tau)$. The maximum is of the order of $\tau^{-3/2}\exp[-(c-2)^2 \tau/4]$. The width of the maximum is
$\mathcal{O}(1)$, and the integrand decays exponentially outside the region $|\chi-x_f(\tau)|\lesssim 1$. Furthermore, the integrand tends to zero exponentially at $\tau \to \infty$ uniformly on any finite interval of $\chi$, and at $\chi\to\pm\infty$ uniformly on any finite interval of $\tau$. As a result, the iterated integral in Eq.~(\ref{formula}) converges (even at $c=2$), and the iterated improper integrals over $\tau$ and over $\chi$ commute, see e.g. Ref.~\cite{10.9}.
At $c\geq 4$ and large $\tau$, the integrand has a maximum at $\chi\simeq c\tau/2$. Its width is of the order of $\sqrt{\tau}$, and the maximum value, up to a pre-exponential factor, is $\sim \exp[\tau\, (1-c^2/8)]$. As $\tau\to\infty$, the maximum goes to zero exponentially fast. Again, this leads to convergence and commutativity of the iterated improper integrals over $\tau$ and over $\chi$.

\subsection{ $c\gg 1$}
\label{Lc}

We derive now an asymptotic expansion of $\phi(c)$ from Eq.~(\ref{formula}) in the inverse powers of $c$.  At large $c$, the dominant contribution to the double integral (\ref{formula}) comes from
small $\tau$. Therefore, we need to expand  $u(x,t)$ at $t\ll 1$. Going over from $x$ and $t$ to $\zeta=x/\sqrt{4t}$ and $t$, one can rewrite the FKPP equation for $u(x,t)=U(\zeta,t)$ as
\begin{equation}\label{FKPP2}
\partial_{\zeta}^2 U+ 2 \zeta \partial_\zeta U=4 t \,(\partial_tU-U+U^2).
\end{equation}
This equation needs to be solved with the initial condition $U(\zeta,0)=\theta(-\zeta)$ and the boundary conditions $U(-\infty,t)=1$ and  $U(\infty,t)=0$. We look for the solution in the form of a power series in $t$:
\begin{equation}\label{shorttime}
U(\zeta,t)= U_{0}(\zeta)+t\, U_{1}(\zeta)+t^2\, U_{2}(\zeta)\dots.
\end{equation}
In the zeroth order in $t$ we obtain
$$
U_0^{\prime\prime}+ 2 z U_0^{\prime}=0,
$$
and the proper solution is
\begin{equation}\label{u0sol}
U_0(\zeta)=\frac{1}{2}\, \text{erfc}\,\zeta .
\end{equation}

In the first order in $t$ we obtain an inhomogeneous linear equation, where the forcing term comes from
the previous iteration:
\begin{equation}\label{ODEu1}
U_1^{\prime\prime}+2 \zeta U_1^{\prime} -4 U_1= 4 (U_0^2-U_0) = \text{erf}^2\,\zeta-1.
\end{equation}
Solving it with zero boundary conditions at $\pm \infty$, we obtain
\begin{equation}\label{u1sol}
U_1(\zeta)=
  \frac{\zeta^2}{2} \left(\text{erf}^2\,\zeta-1\right)
   +\frac{\zeta  e^{-\zeta^2}  \text{erf}\,\zeta}{\sqrt{\pi}}
   +\frac{e^{-2 \zeta^2}}{2\pi}.
\end{equation}

In the second order in $t$ we again obtain a linear equation with a forcing term coming from the previous iterations:
\begin{equation}\label{ODEu2}
U_2^{\prime\prime}+2 \zeta U_2^{\prime} -8 U_2= 4 U_1(2U_0-1),
\end{equation}
etc.

Once the functions $U_0(\zeta)$, $U_1(\zeta)$, $\dots$ are known, we can  calculate the function $\phi(c)$ from
Eq.~(\ref{formula}). Let us define the function $w(\zeta,t)= e^{-t}  U^2(\zeta,t)$, where $U(\zeta,t)$
is given by the expansion~(\ref{shorttime}). The function $w(\zeta,\tau)$ absorbs the factor $e^{-\tau}$ which appears
in the integrand of  Eq.~(\ref{formula}). Clearly, $w(\zeta,t)$ has an expansion of the form
\begin{equation}\label{vexpansion}
w(\zeta,t) = w_0(\zeta) + t \ w_1(\zeta) +  \dots + t^n \  w_n(\zeta) +  \dots .
\end{equation}
Here $w_0(\zeta)=U_0^2(\zeta)$, $w_1(\zeta)=2U_0(\zeta)U_1(\zeta)-U_0^2(\zeta)$, etc.
Plugging this expansion into Eq.~(\ref{formula}) and performing simple changes of variables, we obtain
\begin{equation}\label{formula2}
\phi(c) =\frac{8}{c^3} \left[ R_0+ \frac{4}{c^2} R_1+\dots + \left(\frac{4}{c^2}\right)^n R_n + \dots \right]
\end{equation}
where the coefficients $R_n$, $n=0,1,\dots$, are given by
\begin{equation}\label{Rn}
R_n=  \int_0^\infty \tau^n \  d \tau \int_{-\infty}^\infty d\chi \ e^{\chi-\tau} \  w_n\left(\frac{\chi}{\sqrt{4 \tau}}\right) .
\end{equation}
Here we will only calculate the first two coefficients, $R_0$ and $R_1$. They suffice for evaluating $\Phi(c)$ up to,
and including, $1/c^5$.

\subsubsection{Calculation of $R_0$}

We have
\begin{eqnarray}
  R_0 &=& \int_0^\infty  \  d \tau \int_{-\infty}^\infty d\chi \ e^{\chi-\tau} \  U_0^2\left(\frac{\chi}{\sqrt{4\tau}}\right) \nonumber \\
   &=& \frac{1}{4} \int_0^\infty  \  d \tau \int_{-\infty}^\infty d\chi \ e^{\chi-\tau} \  \text{erfc}^2\left(\frac{\chi}{\sqrt{4\tau}}\right) \nonumber \\
   &=& \frac{1}{2} \int_0^\infty  \  d \tau \sqrt{\tau} e^{-\tau} \int_{-\infty}^\infty d\zeta \ e^{2\sqrt{\tau}\, \zeta} \  \text{erfc}^2 \,\zeta .\label{R0}
\end{eqnarray}
The integral over $z$ can be evaluated as follows \cite{Prudnikov}:
\begin{equation}\label{tab}
\int_{-\infty}^{\infty} d\zeta \,e^{2 \sqrt{\tau}\, \zeta}   \ \text{erfc}^2 \,\zeta = \frac{2 e^{\tau}}{\sqrt{\tau}}
\, \text{erfc}\,\left(\sqrt{\frac{\tau}{2}}\right).
\end{equation}
The resulting integral over $\tau$ yields $R_0=1$.

%\vspace{0.5cm}

\subsubsection{Calculation of $R_1$}

Here we need to evaluate the  integral
\begin{equation}
  R_1 = \int_0^\infty  \  d \tau \, \int_{-\infty}^\infty d\chi \ \tau \, e^{\chi-\tau} \,w_1\left(\frac{\chi}{\sqrt{4 \tau}}\right),
  \label{R1a}
\end{equation}
where $w_1(\zeta)=2U_0(\zeta) U_1(\zeta)  -U_0^2(\zeta)$. A numerical evaluation gives $R_1=-0.213\dots$.

%\vspace{0.5cm}

\subsubsection{$\phi(c)$ and $\Phi(c)$ at $c\gg 1$}

\begin{figure}[ht]
\includegraphics[width=0.4\textwidth,clip=]{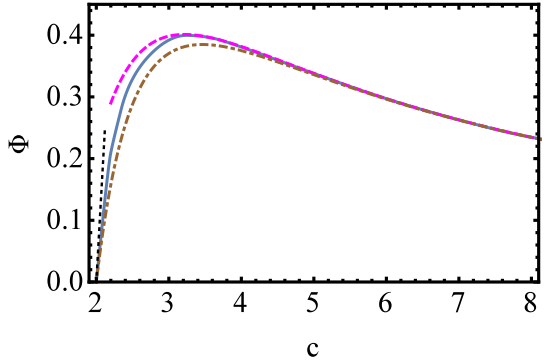}
\caption{Shown is the function $\Phi(c)$, evaluated by a numerical integration of the FKPP equation (the solid line). Also shown are the $c\gg 1$ asymptotic (\ref{Phi5}) (the dashed line) and the $c-2\ll 1$ asymptotic (\ref{Ap170}) (the dotted line).
The dash-dotted line depicts the
lower-order large-$c$ asymptotic (\ref{morecrude}).}
\label{Phi(c)}
\end{figure}

Using Eqs.~(\ref{Phiphi}) and~(\ref{formula2}), we obtain
\begin{equation}\label{phi5}
\phi(c)=\frac{8}{c^3}-
\frac{6.818 \dots}{c^5} +\dots,
\end{equation}
and
\begin{equation}\label{Phi5}
\Phi(c) = \frac{2}{c}-\frac{8}{c^3}+
\frac{6.818 \dots}{c^5} +\dots .
\end{equation}
Figure \ref{Phi(c)} shows the function $\Phi(c)$ evaluated numerically (the solid line), and the asymptotic (\ref{Phi5}) (the dashed line). Also shown by the dash-dotted line is the
lower-order asymptotic
\begin{equation}\label{morecrude}
\Phi(c)\simeq  \frac{2}{c}-\frac{8}{c^3}
\end{equation}
which, by chance, correctly vanishes at $c=2$.  (That the function $\Phi(c)$  indeed vanishes at $c=2$ is shown in the next section.)

%\vspace{0.5cm}
\subsection{$c-2\ll 1$}
\label{Sc-2}

In this subsection we evaluate $\phi(c)$ from  Eq.~(\ref{formula}) for $c-2\ll 1$.
In order to deal with small $\Delta c\equiv c-2$, it is convenient to rewrite Eq.~(\ref{formula}) in the following identical form:
\begin{equation}\label{Ap008}
\phi(c)= \int\limits_{0}^{\infty} d\tau\, \int\limits_{-\infty}^{\infty}d\chi\, e^{\frac{\Delta c}{2}(\chi-2\tau)-\frac{\Delta c^2}{4}\tau}\, e^{\chi-2\tau}u^2(\chi,\tau).
\end{equation}

Let us first calculate
\begin{equation}\label{Ap010}
\phi(2)= \int_{0}^{\infty} \!\!d\tau\int_{-\infty}^{\infty} \!\! d\chi\,\,
 e^{\chi-2 \tau}
 u^2(\chi,\tau).
\end{equation}
By virtue of Eq.~(\ref{FKPP})
\begin{equation}\label{Ap020}
e^{x-2 t} u^2=\partial_x^2\left(e^{x-2 t}\, u\right)-\left(2\partial_x+\partial_t\right)\left(e^{x-2 t}\, u \right)\, .
\end{equation}
Plugging this expression into Eq.~(\ref{Ap010}), we observe that the integral is determined by the behavior of the solution $u(x,t)$ at the boundaries $x=\pm\infty$, $t=0$ and $t\to+\infty$. Indeed,
\begin{eqnarray}
  \phi(2)&=&
  \int_{0}^{\infty} \!\! \left[\partial_\chi\left(e^{\chi-2 \tau}\, u(\chi,\tau)\right)\right]\Bigr|_{\chi=-\infty}^{\chi=+\infty}\, d\tau
 \nonumber \\
&-&
  2\int_{0}^{\infty} \!\! \left(e^{\chi-2 \tau}\, u(\chi,\tau)\right)\Bigr|_{\chi=-\infty}^{\chi=+\infty}\, d\tau
\nonumber \\
&-&
\int\limits_{0}^{+\infty} d\tau \int\limits_{-\infty}^{+\infty}d\chi\,\partial_\tau\left[e^{\chi-2 \tau}\, u(\chi,\tau) \right] \,
  \label{Ap022}.
\end{eqnarray}
The first two terms vanish because $u(x,t)$ tends to 1 at $x\to -\infty$, and decays faster than $e^{-x}$ at $x\to \infty$. To evaluate the third term,  let us rewrite it as
\begin{equation}\label{Ap011}
-\lim\limits_{T\to\infty}\int\limits_{0}^T d\tau\, \int\limits_{-\infty}^{+\infty}d\chi\,\partial_\tau\left[e^{\chi-2 \tau}\, u(\chi,\tau) \right].
\end{equation}
The improper integral over $\chi$, as a function of $\tau$, converges uniformly on any interval $0<\tau<T$. Therefore, the integrals over $\tau$ and over $\chi$ commute~\cite{10.8},
and we obtain
\begin{equation}\label{Ap012}
\phi(2)=J_0-J_{\infty}\, ,
\end{equation}
where
\begin{eqnarray}
  J_0 &=& \int_{-\infty}^{\infty} \!\!\left(e^{\chi-2 \tau}\, u(\chi,\tau) \right) \Bigr|_{\tau=0}\, d\chi \nonumber \\
     &=& \int_{-\infty}^{+\infty} \!\! u(\chi,0)\, e^{\chi}\, d\chi=\int_{-\infty}^{0} \!\! e^{\chi}\, d\chi = 1\, , \label{Ap014}
\end{eqnarray}
and
 \begin{equation}\label{Ap016}
J_{\infty}=\lim\limits_{\tau\to +\infty}\int_{-\infty}^{\infty} \!\! {e^{\chi-2 \tau}}\, u(\chi,\tau)\, d\chi\, .
\end{equation}
To evaluate the integral, entering Eq.~(\ref{Ap016}), at $\tau\gg1$, we use a simplified form of the asymptotic~(\ref{A140}) at $t\gg1$ and $\ln t\ll x-2 t \sim \sqrt{t}$: $u(x,t)\simeq \alpha t^{-3/2}(x-2t)\exp(t-x^2/4t)$ \footnote{Here we use the asymptotic $u(x,t)\simeq \alpha\,t^{-3/2}(x-2t)\exp(t-x^2/4t)$ only in the region  $\ln t\ll x-2t \sim \sqrt{t}$ which dominates the integral $J_{\infty}$. As we will show shortly, the applicability domain of this asymptotic is actually much broader: $\ln t \ll x-2t\ll t$.}. This yields, at $\tau\gg 1$,
\begin{eqnarray}
  &&\int_{-\infty}^{\infty} \!\! {e^{\chi-2 \tau}}\, u(\chi,\tau)\, d\chi =\alpha\, \tau^{-3/2} \int_0^{\infty} \!\! s\, e^{-s^2/4\tau}\, ds\nonumber \\
   &+&{\cal O}(1/\tau)=2\alpha\, \tau^{-1/2}+{\cal O}(1/\tau)\, ,
   \label{Ap018}
\end{eqnarray}
where $s=\chi-2\tau$. Hence $J_{\infty}=0$, and
\begin{equation}\label{phi(2)}
\phi(2)= 1\mbox{~~and~~} \Phi(2)=0
\end{equation}
[see Eq.~(\ref{Phiphi})].
Now we evaluate the leading term of the asymptotic expansion of the difference $\phi(c)-\phi(2)$ in the small parameter $\Delta c$.  $\phi(c)$ and $\phi(2)$ are defined by Eqs.~(\ref{Ap008}) and~(\ref{Ap010}), respectively. Their difference, therefore,  can be written as a sum of the following three integrals:
\begin{equation}\label{Ap92}
\phi(c)-\phi(2)=J_1+J_2+J_3\, ,
\end{equation}
where
\begin{equation}
J_1 = \int\limits_{0}^{\infty} d\tau\, \int\limits_{-\infty}^{\infty} d\chi\,\left[e^{\frac{\Delta c}{2}(\chi-2\tau)}-1\right]\, e^{\chi-2\tau}\, u^2(\chi,\tau) ,  \label{AP94}
\end{equation}
\begin{equation}
J_2 = \int\limits_{0}^{\infty} d\tau\, \int\limits_{-\infty}^{\infty} d\chi\, \left(e^{-\frac{\Delta c^2}{4}\tau} -1\right) e^{\chi-2\tau}u^2(\chi,\tau), \label{AP96}
\end{equation}
and
\begin{eqnarray}
 J_3 &=& \int\limits_{0}^{\infty} d\tau\, \int\limits_{-\infty}^{\infty}d\chi\,\left\{\left(e^{-\frac{\Delta c^2}{4}\tau} -1\right) \right.\nonumber \\
  &\times& \, \left.\left[e^{\frac{\Delta c}{2}(\chi-2\tau)}-1\right]\, e^{\chi-2\tau}u^2(\chi,\tau)\right\} \,.\label{AP98}
\end{eqnarray}
These integrals are evaluated at $\Delta c\to 0$ in Appendix~\ref{App_A}, and the results are:
\begin{eqnarray}
  J_1&=& -\frac{1}{2}\Delta c -\sqrt{\pi}\, \alpha\, \Delta c+{\mathcal O}(\Delta c^2)\, , \label{AP300}\\
 J_2&=&-\sqrt{\pi}\, \alpha\, |\Delta c|+{\mathcal O}(\Delta c^2)\, , \label{AP310}\\
 J_3&=&\Delta c\, |\Delta c|\,{\mathcal O}\left(\bigl|\ln |\Delta c|\bigr|\right)\, .\label{AP320}
\end{eqnarray}
To remind the reader, $\alpha\simeq 0.59$, see Eq.~(\ref{A140}). Substituting Eqs.~(\ref{AP300}) - (\ref{AP320}) into Eq.~(\ref{Ap92}) we obtain, for $|\Delta c|\ll 1$:
\begin{equation}\label{Ap160}
\phi(c)=1-\frac{\Delta c}{2}-2\sqrt{\pi}\, \alpha\, \Delta c\, \theta( \Delta c) +o\left(|\Delta c|^{2-\varepsilon}\right)\,,
\end{equation}
where $\theta(\dots)$ is the step function, and $\varepsilon$ is an arbitrary positive number. As one can see, the integral $J_3$ does not contribute to the leading order. Interestingly, $\phi(c)$ does not have a regular derivative at $c=2$. Plugging Eq.~(\ref{Ap160}) into Eq.~(\ref{Phiphi}), we obtain the following asymptotic of $\Phi(c)$ at $ 0< c-2 \ll 1$:
\begin{equation}\label{Ap170}
\Phi(c)\simeq 2\sqrt{\pi}\, \alpha\, (c-2)\, .
\end{equation}
With this $\Phi(c)$, Eq.~(\ref{Phiphi}) coincides with Eq.~(\ref{A140}), if we substitute  in the latter $x=ct$ and neglect the relatively small $\ln t$ term. The straight-line asymptotic (\ref{Ap170}) is shown in Fig.~\ref{Phi(c)}.

Now we can try to determine  the applicability domain of the asymptotic (\ref{Ap170}).
For $c-2 \ll1$ the integrals $J_2$ and $J_3$ are dominated by $\chi \simeq 2 \tau$ and $\tau \sim (c-2)^{-2}$, see  Appendix~\ref{App_A}. As in the derivation of (\ref{formula}) we used the fact that $\tau \ll t$,
we must demand that
\begin{equation}\label{appl2}
(c-2)^2\,t\gg 1\, .
\end{equation}
As one can check, once this strong inequality holds, the simplifications (i) and (ii) after Eq.~(\ref{C010}) are also legitimate. The criterion (\ref{appl2}), alongside with $c-2\ll 1$, is equivalent to the double strong inequality
\begin{equation}\label{doublestrong}
\sqrt{t}\ll x-2t\ll t\,.
\end{equation}
As a result, the asymptotic~(\ref{Ap170}) has a joint domain of applicability with the asymptotic~(\ref{A140}). The right strong inequality in Eq.~(\ref{doublestrong})
includes $t$ rather than $t/\ln t$ as in the criterion~(\ref{rightinequality}). Essentially, we have extended the applicability domain of  the asymptotic~(\ref{A140}), so that it holds at
\begin{equation}\label{extended}
1 \ll x-2t+\frac{3}{2}\ln t \ll t.
\end{equation}

To summarize this section, the long-time leading-edge asymptotic of $u(ct,t)$ has the form of
Eq.~(\ref{formula1}), where $\Phi(c)=2/c-\phi(c)$, and $\phi(c)$ is determined by Eq.~(\ref{formula}). The plot of $\Phi(c)$ is
depicted in Fig. \ref{Phi(c)}.  Equation~(\ref{formula1}) is valid when $t\gg 1$ and $(c-2)^2\,t \gg 1$. Finally, the asymptotic expansion of $\Phi(c)$ in the powers of $1/c$ is given by Eqs.~(\ref{Phi5}), whereas the leading term of the expansion in the small parameter $0<c-2\ll 1$  is given by Eq.~(\ref{Ap170}).

\section{Evaluation of ${\mathcal P}(c,t)$}
\label{probdens}

\subsection{One seed particle}

Now we use Eqs.~(\ref{E050}), (\ref{formula1}), and the fact that, for a single seed particle, $P(x,t)=u(x,t)$,
to calculate the probability distribution  ${\mathcal P}(c,t)$ of the empirical velocity $c$  at large $t$. We obtain, in the leading order,
\begin{equation}
{\mathcal P}(c,t)\simeq \frac{c \,\Phi(c)}{4\sqrt{\pi}}\, \sqrt{t}\, \exp\left[-t\left(\frac{c^2}{4}-1\right)\right],
\label{E080}
\end{equation}
in agreement with Eq.~(\ref{result}). The prefactor includes a large factor $t$ and depends on $c$ via $c\,\Phi(c)$. The large-$c$ expansion of this quantity only includes even powers of $1/c$:
\begin{equation}\label{cPhi}
c \,\Phi(c)=2 -\frac{8}{c^2}+
\frac{6.818\dots}{c^4}+\dots .
\end{equation}
At $\ln t/t \ll c-2\ll 1$ we can use Eq.~(\ref{Ap170}) and obtain
\begin{equation}
{\mathcal P}(c,t)\simeq \alpha (c-2)\, \sqrt{t}\, \exp\left[-t\left(\frac{c^2}{4}-1\right)\right]\,.
\label{E0800}
\end{equation}

\subsection{More than one seed particle}

Now let the branching Brownian motion start, at $t=0$,  with $M$  seed particles  at positions $x_1=0 \ge x_2 \ge \cdots x_M$. Then from Eq.~(\ref{E040}) one gets for $c>2$
$$
P(ct,t) \simeq \sum_{m=1}^M u(ct-x_m,t).
$$
Using the asymptotic~(\ref{formula1}), we obtain
 at large $t$ and $c>2$
\begin{equation}
P(ct,t)\simeq \frac{e^{-t\,\left(\frac{c^2}{4}-1\right)}}{2\sqrt{\pi t}} \Phi(c)
\sum\limits_{m=0}^M \, e^{\frac{c x_m}{2}}\,.
\label{E150}
\end{equation}
This gives, using Eq.~(\ref{E050}),
\begin{equation}
{\mathcal P}(c,t)\simeq  \frac{\sqrt{t}
 e^{-t\,\left(\frac{c^2}{4}-1\right)}
 }{4\sqrt{\pi}}\,\, c \, \Phi(c)
\sum\limits_{m=0}^M \, e^{\frac{c x_m}{2}}\,.
\label{E170}
\end{equation}
We see that for large $c$ the prefactor is dominated by the position of the rightmost seed.
Expression (\ref{E170}) can also be used  for an infinite number of seed particles, by taking the limit
$M\to \infty$  provided that the density of seed particles does not grow too fast along the negative axis, to keep  the sum over $m$ convergent. Otherwise, even the leading exponential factor in (\ref{E170})  could be changed.

\section{Discussion}
\label{discussion}

In this work we have shown how to calculate, in the long time limit, the prefactor of the probability distribution ${\mathcal P(c,t)}$  of the empirical velocity $c$ of the rightmost particle  of the branching  Brownian motion.
To this end we have studied the long-time asymptotic of the solution of the FKPP equation positions $x=c t$ with $c>2$, when starting from the  initial condition (\ref{step}).  Our main results are Eqs.  (\ref{formula}), (\ref{formula1}),
(\ref{Phiphi}),
and (\ref{E080}) - (\ref{E0800}).

The approach can be generalized to other steep initial conditions.  It was shown in \cite{BD2009} that the solution of the FKPP equation with more general steep intial conditions gives information on the statistics of the positions of the leading particles (for example, the distance between the first and the second particle). It would be interesting to see whether the  method used here allows  one  to calculate, for $c>2$, these statistics,  in particular the distance between the leading particles, conditioned on the position $ct$ of the rightmost one.

The approach can also be extended to other FKPP-like equations. In  Appendix B a generalization to an equation discrete in space and  time is discussed.

%\ \\ \

\section*{ACKNOWLEDGMENTS}
BD thanks the hospitality of the Racah Institute of Physics at the Hebrew University in Jerusalem  where this work was started. BM thanks Ofer Zeitouni for a useful discussion. Financial support for this research was provided in part by grant No.\
2012145 from the United States-Israel Binational Science Foundation (BSF) (BM) and Grant No. 13-01-00314 from the the Russian Foundation for Basic Research (PVS).

\appendix

\section{Evaluation of the integrals $J_1$, $J_2$ and $J_3$ from Eqs.~(\protect\ref{AP94}) - (\protect\ref{AP98})}
\label{App_A}

\subsection{General}
\label{rem}

In this Appendix we evaluate the integrals $J_1$, $J_2$ and $J_3$ from Eqs.~(\ref{AP94}) - (\ref{AP98}) in the leading order in $|\Delta c|$, and estimate the residual terms.

The integrands of $J_1$, $J_2$ and $J_3$  all include the expression $\mu=e^{\chi-2\tau}u(\chi,\tau)^2$ which affects their behavior in an important way. Approximating $\chi-2\tau\simeq\chi-x_f(\tau)-(3/2) \ln \tau$,  we can write
$$
\mu \simeq \tau^{-3/2} e^{\chi-x_f(\tau)} \Psi_2^2[\chi-x_f(\tau)],
$$
where we have used the TFS $\Psi_2$ for $u$. The function $e^{\chi-x_f(\tau)} \Psi_2^2[\chi-x_f(\tau)]$ has a maximum (of order of $1$) inside the region of $|\chi-x_f(\tau)|\sim 1$,
and decays exponentially outside of this region. Therefore, the maximum value of $\mu$ behaves, as a function of $\tau$, as $\tau^{-3/2}$.

\subsection{Evaluation of $J_1$}

At $c$ close to $2$  the
main contribution to $J_1$ comes from the region where $|\chi|\sim\tau\sim 1$.  As a result, $\exp [-\Delta c(\chi-2\tau)/2]$ can be
Taylor-expanded at $\Delta c \ll 1$, and we obtain
\begin{equation}\label{Ap400}
J_1=\frac{\Delta c}{2}\int\limits_{0}^{\infty} d\tau\,  \int\limits_{-\infty}^{\infty}d\chi\,(\chi-2\tau)\, e^{\chi-2\tau}\, u^2(\chi,\tau)
+ {\mathcal O}(\Delta c^2)\, .
\end{equation}
Let us denote
$$
\bar{J_1} =\int\limits_{0}^{\infty} d\tau\,  \int\limits_{-\infty}^{\infty}d\chi\,(\chi-2\tau)\, e^{\chi-2\tau}\, u^2(\chi,\tau)\, .
$$
This integral converges in the region of $\chi \sim \tau \sim 1$ and can be evaluated using the exact equality
\begin{eqnarray}
  &&(x-2t)\, e^{x-2t}\, u^2 = (x-2t)\, \partial_x^2 \left(e^{x-2t}\, u\right) \nonumber \\
  &&\;\;\;\;\;\;\;-\left(2\partial_x+\partial_t\right)\left[(x-2t)\, e^{x-2t}\, u\right]\, . \label{Ap416}
\end{eqnarray}
The calculations are similar to those in Eq.~(\ref{Ap022}):
\begin{eqnarray}
   \bar{J}_1  &=&  -\int\limits_0^{+\infty} d\tau\,\int\limits_{-\infty}^{+\infty} ds \,\partial_\tau \left(s e^sw(s,\tau)\right) \nonumber \\
  &=& -\lim\limits_{T\to+\infty}\int\limits_0^T d\tau \int\limits_{-\infty}^{+\infty} ds \,\partial_\tau  \left[s e^sw(s,\tau)\right] \nonumber\\
  &=& -\lim\limits_{T\to+\infty}\int\limits_{-\infty}^{+\infty} ds\int\limits_0^T d\tau\,\partial_\tau  \left[s e^sw(s,\tau)\right] \nonumber\\
 &=& \lim\limits_{T\to+\infty}\int\limits_{-\infty}^{+\infty} \left[s e^sw(s,\tau)\right]\bigr|_{\tau=T}^{\tau=0}\,\, ds=J_4-J_5, \label{Ap420}
\end{eqnarray}
where $s=\chi-2\tau$, $w(s,\tau)=u(s+2\tau,\tau)$,
\begin{eqnarray}
  J_4 &=& \int_{-\infty}^{+\infty} \!\! \left(s e^sw(s,\tau)\right)\bigr|_{\tau=0}\, ds \nonumber \\
  &=& \int\limits_{-\infty}^{0} s e^{s} \, ds=-1\, , \label{Ap430}
\end{eqnarray}
and
\begin{equation}\label{Ap440}
J_5=\lim\limits_{\tau\to+\infty}\int_{-\infty}^{+\infty} \!\! s e^sw(s,\tau)\, ds\, .
\end{equation}
The change of the order of integration in the third line of  Eq.~(\ref{Ap420}) is justified similarly to how it was done for the third term of Eq.~(\ref{Ap022}).

Let us denote the integral in Eq.~(\ref{Ap440}) as $\bar{J}_5$. As in Eqs.~(\ref{Ap016})  and (\ref{Ap018}),  we can evaluate this integral by using the simplified form of the asymptotic~(\ref{A140}) at  $1\ll\ln t\ll x-x_f(t)\sim \sqrt{t}$: $u(x,t)\simeq \alpha t^{-3/2}(x-2t)\exp(t-x^2/4t)$. As a result, we obtain at $\tau\gg 1$:
\begin{equation}\label{Ap450}
\bar{J}_5\to \alpha\, \tau^{-3/2} \int_0^{\infty} \!\! s^2\, e^{-s^2/4\tau}\, ds=2 \sqrt{\pi}\, \alpha\, .
\end{equation}
Combining all these results, we obtain Eq.~(\ref{AP300}).

\subsection{Evaluation of $J_2$}

Let us now evaluate the integral $J_2$ from Eq.~(\ref{AP96}).
The main contribution to this integral comes from the region $\tau\sim\Delta c^{-2}\gg 1$, where we can use the TFS for $u(x,t)$. This yields
\begin{eqnarray}
 J_2&\simeq& \int\limits_{0}^{\infty} d\tau\, \left(e^{-\frac{\Delta c^2}{4}\tau} -1\right) e^{x_f(\tau)-2\tau}\, \int\limits_{-\infty}^{\infty}d\xi\, e^\xi \Psi_2^2(\xi) \nonumber \\
   &+& {\mathcal O}(\Delta c^2)\,, \label{Ap140}
\end{eqnarray}
where $\xi=\chi-x_f(\tau)$. The integral over $\xi$ can be calculated using the identity $e^\xi \Psi_2^2(\xi)=\left[e^\xi \Psi_2(\xi)\right]^{\prime\prime}$ following from Eq.~(\ref{MFeq}):
$$
\int\limits_{-\infty}^{\infty} e^\xi \Psi_2^2(\xi)\, d\xi=\int\limits_{-\infty}^{\infty} \left(e^\xi \Psi_2\right)^{\prime\prime}\, d\xi=\left(e^\xi \Psi_2\right)^{\prime}\bigr|_{-\infty}^{+\infty}=\gamma e^{\xi_0}\, .
$$
As a result, the integral in Eq.~(\ref{Ap140}) can be rewritten in the leading order as
\begin{eqnarray}
   && \gamma e^{\xi_0}\, \int\limits_{0}^{\infty} \left(e^{-\frac{\Delta c^2}{4}\tau} -1\right)e^{x_f(\tau)-2\tau}\, d\tau \nonumber \\
  &=& \alpha  \int\limits_{0}^{\infty}\left(e^{-\frac{\Delta c^2}{4}\tau} -1\right)\,\tau^{-3/2}\, d\tau \nonumber \\
   &=& -\frac{\alpha |\Delta c|}{2}\, \int\limits_{0}^{\infty}\frac{1-e^{-w}}{w^{3/2}}\, dw=-\sqrt{\pi}\, \alpha\, |\Delta c|\, . \label{Ap150}
\end{eqnarray}
This yields Eq.~(\ref{AP310}).

\subsection{Evaluation of $J_3$}

Finally, we show that the integral $J_3$, defined by Eq.~(\ref{AP98}), does not contribute in the leading order, see Eq.~(\ref{Ap150}).   The main contribution to $J_3$ comes from the region $\tau\sim \Delta c^{-2}\gg 1$. This allows us to estimate this integral by using the TFS in the vicinity of $\chi\sim x_f(\tau)$ and assuming that the argument in the exponent in the brackets is small.  As a result, we can estimate
$$
J_3\sim \Delta c  \int\limits_{0}^{\infty}\left(e^{-\frac{\Delta c^2}{4}\tau} -1\right)\,\frac{\ln \tau}{\tau^{3/2}}\, d\tau\sim  \Delta c\, |\Delta c|\, |\ln |\Delta c||\, ,
$$
bringing us to Eq.~(\ref{AP98}).

\section{A traveling wave equation discrete in  space and time}
In this Appendix we  show how the above calculations  for the FKPP equation (\ref{FKPP}) can be easily extended to more general traveling wave equations. We  take as an example the time- and space-discretized version of (\ref{FKPP})
\begin{eqnarray}
\label{FKPPd}
u_i(t+1)& =& u_i(t)+a_1 \, u_{i-1}(t) + a_{-1} \, u_{i+1}(t)
\\
& &  -(a_1+a_{-1})\,  u_i(t)  + b \,  u_i(t)- b\,  u_i(t)^2\,.
\nonumber
\end{eqnarray}
It corresponds to  a  branching random walk where at each time step, a particle jumps to the right with probability $a_1$, to the left with probability $a_{-1}$ or branches with probability $b$.
Here $u_i(t)$ is the probability that there is at least one  particle at  a position $\ge i$.

If one starts with a single particle at the origin, then
\begin{eqnarray*}
 u_i(0)&=& 1  \ \ \ \ \ \ \ {\rm for} \ \ \ \ i \le 0, \\
  u_i (0) &=& 0  \ \ \ \ \ \ \ {\rm for} \ \ \ \ i \ge 1.
\end{eqnarray*}
Let us  denote by $w_i(t)$ the solution of the linear problem
\begin{eqnarray}
w_i(t+1)& =&w_i(t)+a_1\  w_{i-1} (t) + a_{-1} \  w_{i+1}(t)
\label{lineard}
\\ & &  -(a_1+a_{-1}) \  w_i(t)  + b \  w_i(t)
\nonumber
\end{eqnarray}
with an initial condition localized at the origin:
\begin{eqnarray*}
  w_i(0)  &=& 1  \ \ \ \ \ \ \ {\rm for} \ \ \ \ i = 0 \\
  w_i (0) &=& 0  \ \ \ \ \ \ \ {\rm for} \ \ \ \ i \ne 1 .
\end{eqnarray*}
One can show   that
at any time $t$
\begin{equation}
 \sum_i e^{\lambda   \,  i} \  w_i(t) = e^{t \, h(\lambda)}
 \label{eq1}
\end{equation}
where
$$ e^{ \ h(\lambda)} =1+ a_1 \,  e^{\lambda} + b + a_{-1}  \, e ^{-\lambda} - a_1-a_{-1} $$
(Other jump rates: for example, jumps to next nearest neighbors would change $h(\lambda)$, but what
follows would remain the same.)

From (\ref{eq1}) one can extract the long-time asymptotics of $w_i(t)$ for $i=c t$:
$$w_i(t) \simeq \sqrt{\frac{f''(c)}{2 \pi t}  \ e^{-t \, f(c)}} \ \ \ \ \ {\rm where }  \ \ c=\frac{i}{t}$$
where $f(c)$ can be obtained in a parametric way
$$ f(c) = -h(\lambda) + \lambda h'(\lambda) \ \ \ \ \ ; \  \ \ \ \ c= h'(\lambda)$$
one also has $f'(c)=  \lambda$ and $f''(c) =1/h''(\lambda)$,
so that in a parametric form one has
\begin{equation}
\label{wi-asymptotic}
w_i(t) \simeq \frac{
   e^{t \, (h(\lambda) - \lambda h'(\lambda)) }}{
\sqrt{ 2 \pi \, h''(\lambda) \,  t}}
 \ \ \ \ \ {\rm where }  \ \ c=\frac{i}{t} = h'(\lambda)
\end{equation}

One can then show that,  for a finite $k$ and finite $t_0$
\begin{equation}
w_{i+k}(t-t_0)  \sim e^{-\lambda\, k - h(\lambda)t_0}   \ w_i(t)
\label{eq2}
\end{equation}
(The derivation of (\ref{eq2}) is slightly subtle: changing $i$ to $i+k$ and $t$ to $t-t_0$ implies that $c$ is changed to $c+k/t +t_0 c/t$ and therefore $\lambda$ is changed to $\lambda +(k/t+c t_0/t)/h''(\lambda$).)

Therefore, the solution $v_i(t)$ of the linear problem
\begin{eqnarray}
\label{linear-bis}
v_i(t+1)& = & v_i(t)+a_1\  v_{i-1}(t) + a_{-1} \  v_{i+1}(t)
\\ & & -(a_1+a_{-1}) \  v_i(t)  + b \  v_i(t)
\nonumber
\end{eqnarray}
with  a step initial condition
\begin{eqnarray*}
 v_i(0) &=& 1  \ \ \ \ \ \ \ {\rm for} \ \ \ \ i \le 0 \\
  v_i(0)  &=& 0  \ \ \ \ \ \ \ {\rm for} \ \ \ \ i \ge 1
\end{eqnarray*}
is given by
\begin{equation}
v_i(t)= \sum_{k \ge 0}w_{i+k}(t) \simeq \frac{
e^{t \, (h(\lambda) - \lambda h'(\lambda)) }}
{(1-e^{-\lambda}) \ \sqrt{ 2 \pi \, h''(\lambda) \,  t}}  \
\label{vi-asymptotic}
\end{equation}

Now let us  use the following identity which is a direct generalization of (\ref{exact})
\begin{equation}
 \label{exact-bis}
u_i(t) = v_i(t)- b \sum_{t_0=0}^{t-1}  \sum_j w_{i-j}(t-t_0-1) \, u_j(t_0)^2 \ .
\end{equation}
This  becomes in the long time limit, using (\ref{eq2})
$$u_i(t) = v_i(t)- b \, w_i(t) \sum_{t_0=1}^\infty  \sum_j e^{\lambda \, j - h(\lambda) \, (t_0+1)}\   u_j(t_0)^2 $$
Therefore in the long time limit (\ref{wi-asymptotic},\ref{vi-asymptotic}), one gets
\begin{eqnarray}
&u_{i=c \, t}(t)
& \simeq
  \frac{
 e^{t \, (h(\lambda) - \lambda h'(\lambda)) }}
{\sqrt{ 2 \pi \, h''(\lambda) \,  t}}
\times
\nonumber
\\ && \left[ \frac{ 1}{1 - e^{-\lambda}}
-b
 \sum_{t_0=0}^\infty  \sum_j e^{\lambda \, j - h(\lambda) \, (t_0+1)}  \ u_j(t_0)^2
  \right]
\nonumber
\end{eqnarray}
where $c=h'(\lambda)$.
So one expects that for large $t$:
\begin{equation}
{u_{i=c \, t}(t) \simeq
 \frac{
 e^{t \, (h(\lambda) - \lambda h'(\lambda)) }}
{\sqrt{ 2 \pi \, h''(\lambda) \,  t}}
  \Phi(c)
}
\label{B1}
\end{equation}
where
\begin{equation}
\Phi(c)=
\frac{1}{ 1 - e^{-\lambda}}-b
 \sum_{t_0=0}^\infty  \sum_j e^{\lambda \, j  - h(\lambda) \, (t_0+1)}  \ u_j(t_0)^2
\label{B2}
\end{equation}
This is the generalization to the discrete case of Eqs.~(\ref{formula}), (\ref{formula1}) and (\ref{Phiphi}).
A large $\lambda$ expansion  (and therefore the large $c$ behavior) can  then be  obtained as in (\ref{formula}) from the  contributions of the short times $t_0$ in the sum.

\begin{figure}[ht]
\includegraphics[width=0.4\textwidth]{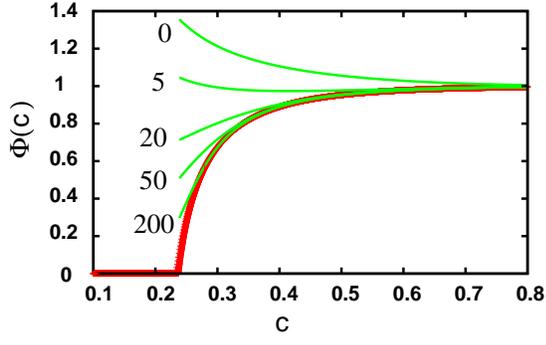}
\caption{
 The function $\Phi(c)$ obtained  from Eq.~(\ref{B1}) by calculating the exact $u_i(t)$ at times $t=1000,2000,\cdots 5000$. The 5 curves are superimposed on the thick curve. The approximations obtained from  our result (\ref{B2}) by truncating the sum overs $t_0$ ($t_0=0, t_0 \leq 5, t_0 \leq 20, t_0 \leq 50, t_0 \leq 200$).}
\label{rbar}
\end{figure}

For an illustration, we have considered the case
$$
a_{-1}=0\,, \ \ \ \  \ \ \ \ \ a_1=b=0.1\,.
$$
For these values the velocity of the front is
$ v_*  \simeq 0.238656...$.
We have measured the function $\Phi(c)$ by numerically calculating the full solution at times 1000, 2000, ... 5000,
and the curves are superimposed to give the thick curve  in Fig. \ref{rbar}.
Then we  have used formula (\ref{B2}) by truncating the sum over $t_0$ to $1,6,21,51,201$ terms.
This gives the thin curves.
The further the truncation is, the better the approximation seems to be.


\begin{thebibliography}{99}
\bibitem{equationnumbers}
The asymptotic (\ref{A010}) corresponds to the case of $\beta=3/2$ in Eqs.~(A2), (A6), (A8) and (A13)  of Ref. \cite{Derridashift}. It is also presented in Eq.~(1.14) of Ref. \cite{EbertSaarloos}.
\end{thebibliography}
\end{document}